\documentclass[prb,preprint,showpacs]{revtex4}

\usepackage{graphicx}
\usepackage{amsmath}

\begin{document}

\title{A simple way of approximating the canonical partition functions in
statistical mechanics}
\author{Francisco M. Fern\'andez}\email{fernande@quimica.unlp.edu.ar}

\affiliation{INIFTA (UNLP, CCT La Plata--CONICET), Blvd.\ 113 y 64
S/N, Sucursal 4, Casilla de Correo 16, 1900 La Plata, Argentina}

\begin{abstract}
We propose a simple pedagogical way of introducing the
Euler-MacLaurin summation formula in an undergraduate course on
statistical mechanics. We put forward two alternative routes: the
first one is the simplest and yields the first two terms of the
expansion. The second one is somewhat more elaborate and takes
into account all the correction terms. We apply both to the
calculation of the simplest one-particle canonical partition
functions for the translational, vibrational and rotational
degrees of freedom.

\end{abstract}


\maketitle

\section{Introduction}

\label{sec:intro}

In statistical mechanics the thermodynamic functions are given in
terms of the logarithm of the partition function and its
derivatives with respect to volume, temperature, etc. The
mathematical expression of the canonical partition function is an
infinite sum over all the states of the system or, under some
simplifying assumptions, the states of the individual
particles.\cite{M73} Some of these infinite sums are commonly
calculated approximately by means of the Euler-MacLaurin summation
formula\cite{AS72} and most textbooks simply show how to apply it
to the cases of interest.\cite{M73} A rigorous derivation of the
summation formula may be rather too demanding for inexperienced
students and one avoids it in introductory courses on statistical
thermodynamics or statistical mechanics.

There is a relatively simple way of deriving the summation formula by means
of operator methods\cite{FC96} but it also requires some kind of
mathematical expertise that students of introductory courses may not posses.
However, the students may feel more confident and confortable if they are
shown how to carry out such calculations by means of mathematical methods
that they have already learned in a first course on mathematical analysis.

In an introductory undergraduate course we show how to obtain the
first two terms of the Euler-MacLaurin summation formula in an
extremely simple way that only requires the students to be
familiar with the Taylor expansion. The aim of this paper is to
put forward this approach that we deem suitable for pedagogical
purposes. In section~\ref{sec:simple_appro} we outline the method
and apply it to some simple examples in
section~\ref{sec:examples}. If we decide to show the students how
to obtain terms of higher order this
simple approach results to be rather cumbersome. For that reason, in section~%
\ref{sec:systematic} we show how to derive the full summation formula by a
relatively minor change of strategy that requires some additional
mathematical skills. The reader may choose one or another approach depending
on the level of the course. Finally, in section~\ref{sec:conclusions} we
summarize the results and draw conclusions.

\section{The simplest approximation}

\label{sec:simple_appro}

The simplest canonical partition functions in statistical
mechanics can be expressed as sums of the form
\begin{equation}
S=\sum_{n=0}^{\infty }f(n),  \label{eq:S}
\end{equation}
where $f(n)$ should tend to zero sufficiently fast when
$n\rightarrow \infty$ because, otherwise, the sum does not
converge. In most cases one cannot obtain this sum in closed form
and therefore resorts to some kind of approximation valid under
certain conditions; for example, sufficiently high temperature. In
order to derive such an approach we define
\begin{equation}
S(x)=\sum_{n=0}^{\infty }f(n+x),  \label{eq:S(x)}
\end{equation}
that satisfies
\begin{eqnarray}
S(0) &=&S  \nonumber \\
\lim\limits_{x\rightarrow \infty }S(x) &=&0,  \label{eq:S(x)_prop1}
\end{eqnarray}
and
\begin{equation}
S(x)-S(x+1)=f(x).  \label{eq:S(x)_prop2}
\end{equation}

Substituting the Taylor expansion $S(x+h)=S(x)+S^{\prime }(x)h+\frac{1}{2}%
S^{\prime \prime }(x)h^{2}+\ldots $ for $h=1$ into (\ref{eq:S(x)_prop2})

we obtain, after some rearrangement,
\begin{equation}
-S^{\prime }(x)=f(x)+\frac{1}{2}S^{\prime \prime }(x)+\ldots .
\end{equation}
If we integrate this expression between $x$ and $\infty $ and take into
account that $S(x)$ and all its derivatives vanish at the upper limit we
have
\begin{equation}
S(x)=\int_{x}^{\infty }f(t)\,dt-\frac{1}{2}S^{\prime }(x)+\ldots ,
\label{eq:S(x)_iter}
\end{equation}
that can be solved iteratively. In the first step we omit the
derivatives of $S(x)$ so that this function is approximately given
by $S(x)$ $\approx \int_{x}^{\infty }f(t)\,dt$. If we substitute
this approximate result into the right-hand side of
(\ref{eq:S(x)_iter}) we have
\begin{equation}
S(x)\approx \int_{x}^{\infty }f(t)\,dt+\frac{1}{2}f(x).
\label{eq:S(x)_approx1}
\end{equation}
Thus, for $x=0$ we obtain a simple approximation to the sum (\ref{eq:S}):
\begin{equation}
S(0)=S\approx \int_{0}^{\infty }f(t)\,dt+\frac{1}{2}f(0).
\label{eq:S_approx_simp}
\end{equation}

The simple method just outlined is suitable for introductory courses because
it only requires basic knowlegde in mathematical analysis. This procedure is
not suitable for the systematic calculation of the corrections of higher
order to the basic formula (\ref{eq:S_approx_simp}) because it soon becomes
rather cumbersome. However, this result is suitable for most porposes in an
introductory course on statistical mechanics because several partitions
functions for atomic and simple molecular systems can be easily derived from
it.\cite{M73}

\section{Some simple examples}

\label{sec:examples}

Under some simplifying assumptions that we will not discuss in
this paper statistical mechanics tell us how to express the
thermodynamic functions of atomic and molecular systems in terms
of one-particle partition functions of the form\cite{M73}
\begin{equation}
q(V,T)=\sum_{n}g_{n}e^{-\epsilon _{n}/(k_{B}T)},  \label{eq:q(V,T)}
\end{equation}
where $V$ is the volume of the container, $T$ the absolute temperature, $%
\epsilon _{n}$ the $n$-th energy level (assumed to be $g_{n}$-fold
degenerate) of the particle and $k_{B}$ the Boltzman constant.

\begin{figure}[H]
\begin{center}
\includegraphics[width=9cm]{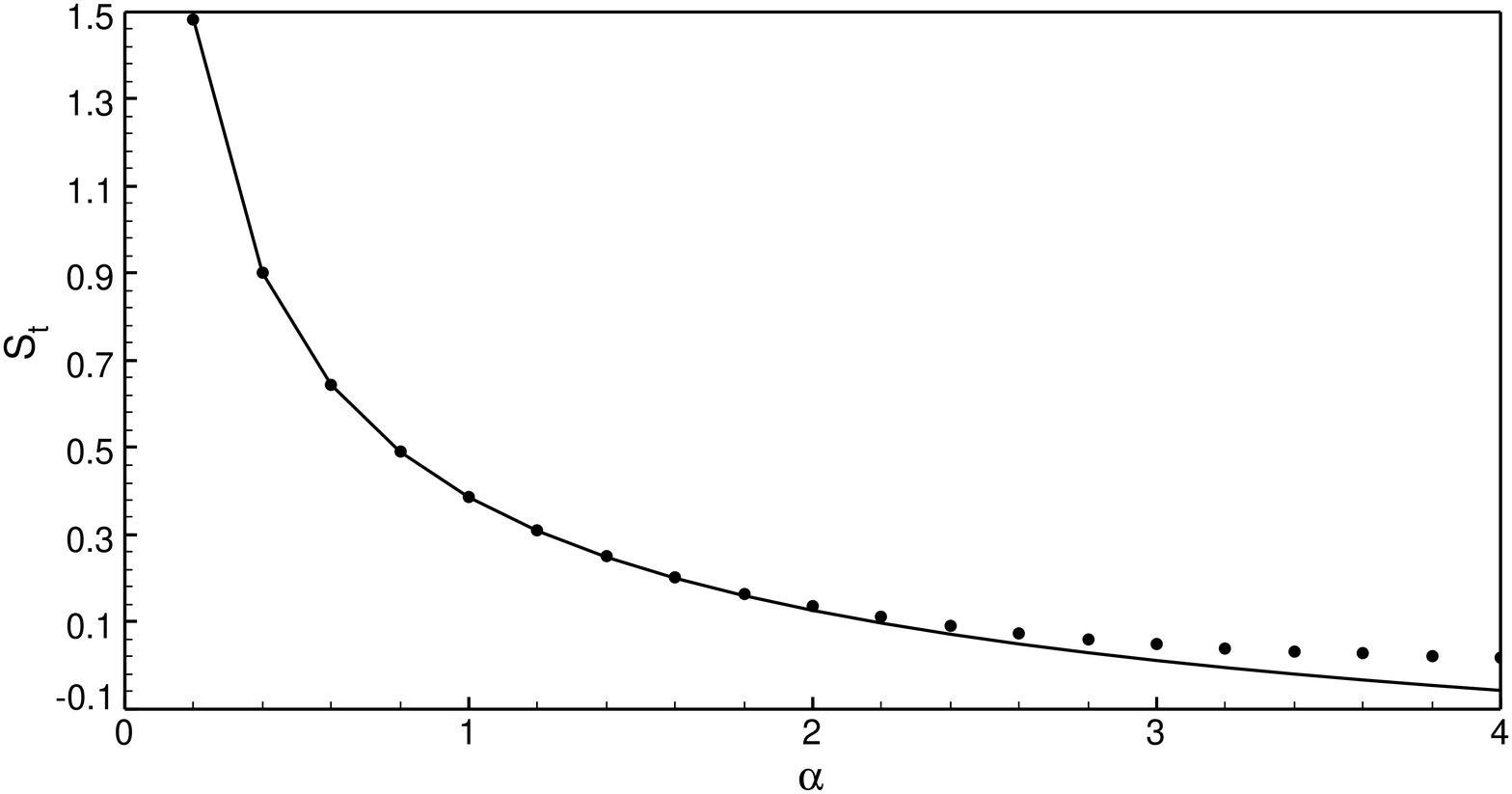}
\end{center}
\caption{$S_t$ calculated by means of the sum (\ref{eq:S_PB}) (points) and
the simple Euler-MacLaurin expression (\ref{eq:S_PB_appro}) (solid line)}
\label{fig:PB}
\end{figure}

Let us first consider a particle of mass $m$ in a one-dimensional box of
length $L$ with impenetrable walls which is the starting point for the
calculation of the thermodynamic properties of an ideal monoatomic gas.\cite
{M73} If the particle has no internal structure its spectrum is only given
by the translational degree of freedom:
\begin{equation}
\epsilon _{n}^{t}=\frac{\hbar^{2}\pi^2
n^{2}}{2mL^{2}},\;n=1,2,\ldots , \label{eq:e_n_PB}
\end{equation}
where $\hbar$ is the Planck constant $h$ divided $2\pi$ and
$g_{n}=1$. Upon defining $\alpha =h^{2}/($ $8mk_{B}TL^{2})$ the
partition function $q_t$ reads
\begin{eqnarray}
q_{t} &=&S_{t}-1  \nonumber \\
S_{t} &=&\sum_{n=0}^{\infty }e^{-\alpha n^{2}},  \label{eq:S_PB}
\end{eqnarray}
and the approximation (\ref{eq:S_approx_simp}) yields
\begin{equation}
S_{t}\approx \frac{\sqrt{\pi }}{2\sqrt{\alpha }}+\frac{1}{2}.
\label{eq:S_PB_appro}
\end{equation}
This approximate expression is accurate for sufficiently small
values of $\alpha $ or, equivalently, when the thermal de Broglie
wavelength $\Lambda =h/\sqrt{2\pi mk_{B}T}$ is much smaller than
the box length $L$.\cite{M73} The addition of corrections of
higher order commonly improves the result of the approximate
partition function, but in this case all of them vanish.\cite{FC96} Figure~%
\ref{fig:PB} shows that this approximate expression becomes increasingly
more accurate as $\alpha $ decreases.

The spectrum of a one-dimensional harmonic oscillator is given by
\begin{equation}
\epsilon _{n}^{v}=\left( n+\frac{1}{2}\right) h\nu ,\;n=0,1,\ldots ,
\label{eq:e_n_HO}
\end{equation}
where $\nu $ is the frequency of the oscillation and $g_{n}=1$. The
partition function for this degree of freedom is given by
\begin{eqnarray}
q_{v} &=&e^{-\alpha /2}S_{v},  \nonumber \\
S_{v} &=&\sum_{n=0}^{\infty }e^{-\alpha n},  \label{eq:S_HO}
\end{eqnarray}
where $\alpha =h\nu /(k_{B}T)$. In this case the sum $S_{v}$ is a
geometric series that can be calculated exactly and therefore we
can derive its small-$\alpha $ expansion in a straightforward way:
\begin{equation}
S_{v}=\frac{1}{1-e^{-\alpha }}=\frac{1}{\alpha }+\frac{1}{2}+\frac{\alpha }{%
12}-\frac{\alpha ^{3}}{720}+\ldots .  \label{eq:S_exact_exp}
\end{equation}
If we define the vibrational temperature $\theta _{v}=h\nu /h_{B}$ then $%
\alpha =\theta _{v}/T$ and the Euler-MacLaurin formula is a good
approximation when $\theta _{v}\ll T$. The approximate expression (\ref
{eq:S_approx_simp}) yields the first two terms of this series exactly.

\begin{figure}[H]
\begin{center}
\includegraphics[width=9cm]{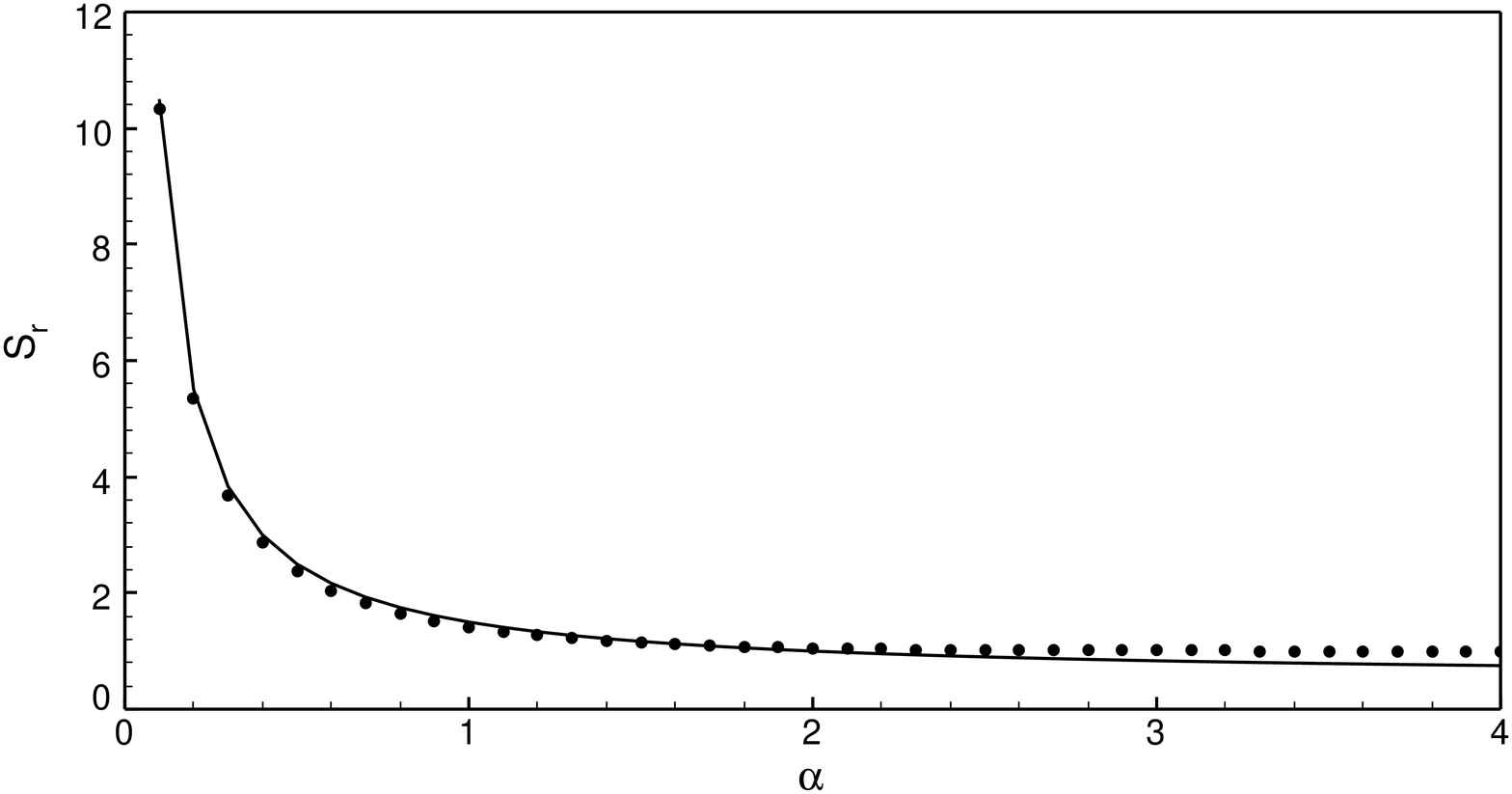}
\end{center}
\caption{$S_r$ calculated by means of the sum (\ref{eq:S_RR}) (points) and
the simple Euler-MacLaurin expression (\ref{eq:S_RR_appro1}) (solid line)}
\label{fig:RR}
\end{figure}

The last example is the partition function for a rigid rotor with moment of
inertia $I$. In this case, every energy level
\begin{equation}
\epsilon _{J}^{r}=\frac{\hbar ^{2}}{2I}J(J+1),\;J=0,1,\ldots ,
\label{eq:e_J_RR}
\end{equation}
is $(2J+1)$-fold degenerate and the partition function for the rotational
degree of freedom is given by
\begin{equation}
q_{r}=S_{r}=\sum_{J=0}^{\infty }(2J+1)e^{-\alpha J(J+1)},  \label{eq:S_RR}
\end{equation}
where $\alpha =\hbar ^{2}/(2Ik_{B}T)$. The approximate expression (\ref
{eq:S_approx_simp}) yields
\begin{equation}
S_{r}\approx \frac{1}{\alpha }+\frac{1}{2},  \label{eq:S_RR_appro1}
\end{equation}
that is accurate enough if $\alpha $ is sufficently small as shown in figure~%
\ref{fig:RR}. It is custommary to define the rotational temperature $\theta
_{r}=\hbar ^{2}/(2Ik_{B})$\cite{M73} so that $\alpha =\theta _{r}/T\ll 1$
when $\theta _{r}\ll T$. It is worth mentioning that the $\alpha $%
-independent term in equation (\ref{eq:S_RR_appro1}) is not exact as shown
in the following section.

\section{Systematic approach}

\label{sec:systematic}

If we decide to show the students how to obtain expression of
higher order it is convenient to proceed in a different way. If we
write the Taylor expansion discussed in
section~\ref{sec:simple_appro} as
\begin{equation}
S(x+1)=\sum_{j=0}^{\infty }\frac{1}{j!}S^{(j)}(x),  \label{eq:S(x+1)_Taylor}
\end{equation}
where $S^{(j)}(x)$ is the $j$-th derivative of $S(x)$ with respect
to $x$, then equation (\ref{eq:S(x)_iter}) reads
\begin{equation}
S(x)=F(x)-\sum_{j=1}^{\infty }\frac{1}{(j+1)!}S^{(j)}(x),
\label{eq:S(x)_iter2}
\end{equation}
where $F(x)=\int_{x}^{\infty }f(t)\,dt$.

Instead of trying to solve equation (\ref{eq:S(x)_iter2}) iteratively we
propose a solution of the form
\begin{equation}
S(x)=\sum_{k=0}^{\infty }a_{k}F^{(k)}(x),  \label{eq:S(x)_ansatz}
\end{equation}
where $F^{(k)}(x)=-f^{(k-1)}(x)$. In order to obtain the coefficients $a_{k}$
we substitute (\ref{eq:S(x)_ansatz}) into (\ref{eq:S(x)_iter2}) and compare
the coefficients of $F^{(n)}(x)$ in the left- and right-hand sides; the
result is
\begin{eqnarray}
a_{n} &=&-\sum_{j=1}^{n}\frac{a_{n-j}}{(j+1)!},\;n=1,2,\ldots ,  \nonumber \\
a_{0} &=&1.  \label{eq:a_n_rec}
\end{eqnarray}
We thus have
\begin{equation}
S(x)=F(x)-\sum_{k=0}^{\infty }a_{k+1}f^{(k)}(x),  \label{eq:S(x)_appro_gen}
\end{equation}
and
\begin{equation}
S=\int_{0}^{\infty }f(t)\,dt-\sum_{k=0}^{\infty }a_{k+1}f^{(k)}(0).
\label{eq:S_appro_gen1}
\end{equation}

Straightforward inspection of the first coefficients $a_{n}$
\begin{eqnarray}
a_{1} &=&-\frac{1}{2},\,a_{2}=\frac{1}{12},\,a_{3}=0,\,a_{4}=-\frac{1}{720}%
,\,a_{5}=0,\,a_{6}=\frac{1}{30240},\,a_{7}=0,\,  \nonumber \\
a_{8} &=&-\frac{1}{1209600},\,a_{9}=0,\,a_{10}=\frac{1}{47900160},
\label{eq:a_n}
\end{eqnarray}
suggests that $a_{2j+1}$ vanish for all $j>0$. In order to prove this
conjecture we consider the function
\begin{equation}
u(x)=\frac{x}{e^{x}-1},  \label{eq:u(x)}
\end{equation}
that satisfies
\begin{equation}
u(x)-u(-x)=-x.  \label{eq:u(x)-u(-x)}
\end{equation}
If we substitute the Taylor expansion
\begin{equation}
u(x)=\sum_{j=0}^{\infty }u_{j}x^{j},  \label{eq:u(x)_series}
\end{equation}
into equation (\ref{eq:u(x)-u(-x)}) we conclude that $u_{2j+1}=0$
for all $j>0$. If we now expand $(e^{x}-1)u(x)=x$ in a Taylor
series about $x=0$ and compare the coefficients of $x^{n}$ in the
left- and right-hand sides of the resulting equation we obtain a
recurrence relation for the coefficients $u_{n}$ that is identical
with equation (\ref{eq:a_n_rec}) for the coefficients $a_{n}$. We
thus conclude that $u_{n}=a_{n}$ for all $n$. Therefore, equation
(\ref{eq:S_appro_gen1}) reduces to
\begin{equation}
S=\int_{0}^{\infty }f(t)\,dt+\frac{1}{2}f(0)-\sum_{k=1}^{\infty
}a_{2k}f^{(2k-1)}(0).  \label{eq:S_appro_gen2}
\end{equation}
It is worth noting that the coefficients $a_{j}$ are related to the
Brillouin numbers $B_{j}$\cite{AS72} in the following way: $a_{j}=B_{j}/j!$.

The operator method leads to this result in a more straightforward way and
is also more convenient for the discussion of the radius of convergence of
the series.\cite{FC96} However, we do not discuss it here because it
requires the introduction of functions of operators that may not be suitable
for an undergraduate course.

Let us apply this more accurate approach to the examples discussed in
section~\ref{sec:examples}. As pointed out in that section nothing can be
done with the particle in a box because all the corrections of higher order
vanish. This surprising result can be explained very easily and applies to
all the sums in which $f(-x)=f(x)$ because $f^{(2k-1)}(-x)=-f^{(2k-1)}(x)$
and $f^{(2k-1)}(0)=0$. A more rigorous analysis of such problems, based on
the Poisson summation formula, can be found eslewhere.\cite{FC96}

The harmonic oscillator is a suitable simple example for testing
the summation formula (\ref{eq:S_appro_gen2}) because we can
calculate the exact expansion as shown in equation
(\ref{eq:S_exact_exp}). In this case every new term added to the
summation formula (\ref{eq:S_appro_gen2}) yields one more term of
the small-$\alpha $ series (\ref{eq:S_exact_exp}) as one may
easily verify.

The application of the summation formula (\ref{eq:S_appro_gen2})
to the partition function for the rigid rotor should be carried
out with care. The reason is that $f^{(2k+1)}(0)=\alpha
^{k}P_{k+1}(\alpha )$, where $P_{k+1}(\alpha )$ is a polynomial
function of $\alpha $ of degree $k+1$ and $P_{k+1}(0)\neq 0$ .
Therefore, the summation formula that includes all the terms
through $f^{(2k+1)}(0)$ only yields the small-$\alpha $ series
correctly through degree $k$. For this reason the simple
expression (\ref{eq:S_approx_simp}) does not yield the correct
$\alpha $-independent term that receives
contributions from $f(0)$ and $f^{\prime }(0)$. The expansion accurate to $%
\alpha ^{4}$ is easily shown to be
\begin{equation}
S_{r}=\frac{1}{\alpha }+\frac{1}{3}+\frac{\alpha }{15}+\frac{4\alpha ^{2}}{%
315}+\frac{\alpha ^{3}}{315}+\frac{4\alpha ^{4}}{3465}+\ldots ,
\label{eq:S_RR_EM}
\end{equation}
if we add all the terms trough $f^{(9)}(0)$ in the summation
formula (\ref {eq:S_appro_gen2}) as argued above. Since this
series does not converge\cite {FC96} one should truncate it before
the terms start to increase.

\section{Conclusions}

\label{sec:conclusions}

As stated in the introduction we show the students the simple
method developed in section~\ref{sec:simple_appro} because it is
sufficient for the purposes of our course. However, if a motivated
student wants to learn how to obtain the corrections of greater
order that appear in some of the available texbooks on statistical
mechanics then one can suggest him or her to try the systematic
approach of section~\ref{sec:systematic} or even the operator
method.\cite{FC96} In our opinion the present way of deriving the
Euler-MacLaurin summation formula for statistical mechanics
applications is more convenient for pedagogical purposes than the
traditional one that appears in most textbooks on mathematics or
numerical analysis.\cite{A99} Such traditional approaches are
certainly more rigorous and general but require the students to be
more experienced in mathematics.

In passing, it is worth mentioning that the summation formula
given by equations (\ref{eq:a_n_rec}) and\ (\ref{eq:S_appro_gen2})
is suitable for motivating the students to resort to a computer
algebra system. In this way they bypass the tedious algebraic
manipulation of the equations that is necessary for the
calculation of contributions of large order like those in equation
(\ref{eq:S_RR_EM}) and practise programming in any of such useful
languages.

\end{document}